\documentclass[12pt,letterpaper]{article}
\usepackage{jheppub}
\usepackage[utf8x]{inputenc}
\pdfoutput=1
\usepackage{graphicx}
\usepackage{subfigure}
\usepackage{amsmath}
\usepackage{bm} 
\usepackage{xspace}
\usepackage{relsize}
\usepackage{placeins}
\usepackage{multirow}
\usepackage{soul}
\usepackage{ulem}

\newcommand{\nc}{\newcommand}

\usepackage{slashed}

\nc{\beq}{\begin{equation}}
\nc{\eeq}{\end{equation}}
\nc{\beqa}{\begin{eqnarray}}
\nc{\eeqa}{\end{eqnarray}}
\nc{\bea}{\begin{eqnarray}}
\nc{\eea}{\end{eqnarray}}
\nc{\barray}{\begin{eqnarray}}
\nc{\earray}{\end{eqnarray}}
\nc{\barrayn}{\begin{eqnarray*}}
\nc{\earrayn}{\end{eqnarray*}}
\nc{\ra}{\rightarrow}
\newcommand{\lsim}{\!\mathrel{\hbox{\rlap{\lower.55ex \hbox{$\sim$}} \kern-.34em \raise.4ex \hbox{$<$}}}}
\newcommand{\gsim}{\!\mathrel{\hbox{\rlap{\lower.55ex \hbox{$\sim$}} \kern-.34em \raise.4ex \hbox{$>$}}}}
\nc{\Tr}{{\rm Tr}}
\nc{\slsh}{\slash\hspace*{-0.22cm}}

\def\be{\begin{equation}}
\def\ee{\end{equation}}
\def\bea{\begin{eqnarray}}
\def\eea{\end{eqnarray}}

\nc{\infinity}{\infty}
\nc{\mc}{\mathcal}
\nc{\M}{\mathcal{M}}

\def\to{\rightarrow}

\allowdisplaybreaks

\begin{document}

\title{Explaining \boldsymbol{$\mathbf (g-2)_\mu$} with Multi-TeV Sleptons}

\author[a]{Wolfgang Altmannshofer,}
\author[a]{Sri Aditya Gadam,}
\author[a]{Stefania Gori,}
\author[a]{Nick Hamer}

\affiliation[a]{Santa Cruz Institute for Particle Physics and Department of Physics, University of California, Santa Cruz, 1156 High Street, Santa Cruz, CA 95064, USA}

\emailAdd{waltmann@ucsc.edu}
\emailAdd{sgori@ucsc.edu}
\emailAdd{sgadam@ucsc.edu}
\emailAdd{nhamer@ucsc.edu}

\abstract{We present a supersymmetric extension of the Standard Model in which the new physics contributions to the anomalous magnetic moment of the muon can be more than an order of magnitude larger than in the minimal supersymmetric Standard Model. The extended electroweak symmetry breaking sector of the model can consistently accommodate Higgs bosons and Higgsinos with $O(1)$ couplings to muons. We find that sleptons with masses in the multi-TeV range can comfortably explain the recently confirmed discrepancy in the anomalous magnetic moment of the muon. We discuss additional phenomenological aspects of the model, including its effects on tau flavor changing decays.
}

\maketitle

\section{Introduction}

The Higgs discovery and the subsequent measurements of its properties at the LHC have been a crucial confirmation of the Standard Model (SM) of particle physics. Using data from Run I and Run II of the LHC, we know that the Higgs has SM-like properties and that its couplings to SM gauge bosons and third generation fermions agree with the SM predictions at the $10\%-20\%$ level~\cite{Aad:2019mbh,Sirunyan:2018koj}. Much less is known about the Higgs couplings to first and second generation fermions. Only recently, the LHC showed the first evidence for the Higgs coupling to muons \cite{Aad:2020xfq, Sirunyan:2020two}. There is no evidence yet for the Higgs couplings to light quarks and electrons. 

At the same time, the origin of the large hierarchies in the SM fermion masses, as well as the hierarchical structure of the CKM quark mixing matrix, constitute a long-standing open question in particle physics: the so-called SM flavor puzzle. One dynamical approach to this puzzle is to couple the first two generations exclusively to a new subleading source of electroweak symmetry breaking, in the form of a second Higgs doublet or some strong dynamics~\cite{Altmannshofer:2015esa, Ghosh:2015gpa, Botella:2016krk} (see also~\cite{Das:1995df, Blechman:2010cs, Egana-Ugrinovic:2019dqu} for related ideas).
Such a scenario adds new sources of flavor universality breaking to the SM. This is an experimentally viable option due to our lack of knowledge of the Higgs couplings to first and second generations. One concrete realization of this scenario is the Flavorful-Two-Higgs-Doublet-Model (F2HDM)~\cite{Altmannshofer:2016zrn}.

Recently, new measurements involving muons have been reported by precision experiments. The LHCb
collaboration has reported updated results on the ratio $R_K$ of the rare $B$ meson decay rates $B\to K\mu\mu$ and $B\to Kee$~\cite{Aaij:2021vac}, confirming earlier hints~\cite{Aaij:2014ora, Aaij:2017vbb, Aaij:2019wad} for lepton flavor universality violation in rare $B$ decays.
In addition, a new measurement of the muon anomalous magnetic moment $a_\mu\equiv (g_\mu-2)/2$ has been very recently reported by the Fermilab Muon g-2 collaboration, $a_\mu^{\rm{FNAL}}=116~592~040(54)\times 10^{-11}$ \cite{Abi:2021gix}. This measurement is consistent with the previous measurement by the E821 experiment at BNL, $a_\mu^{\rm{BNL}}=116~592~089(63)\times 10^{-11}$~\cite{Bennett:2006fi}. The quoted combination of experimental results is $a_\mu^{\rm{exp}}=116~592~061(41)\times 10^{-11}$ and deviates by 4.2$\sigma$ from the SM prediction reported by the white paper of the g-2 theory initiative, $a_\mu^{\rm{SM}}=116~591~810(43)\times 10^{-11}$~\cite{Aoyama:2020ynm}, based on~\cite{Davier:2017zfy, Keshavarzi:2018mgv, colangelo:2018mtw, hoferichter:2019gzf, davier:2019can, keshavarzi:2019abf, kurz:2014wya,  chakraborty:2017tqp, borsanyi:2017zdw, blum:2018mom, giusti:2019xct, shintani:2019wai, Davies:2019efs, gerardin:2019rua, Aubin:2019usy, giusti:2019hkz, melnikov:2003xd, masjuan:2017tvw, Colangelo:2017fiz, hoferichter:2018kwz, gerardin:2019vio, bijnens:2019ghy, colangelo:2019uex, pauk:2014rta, danilkin:2016hnh, jegerlehner:2017gek, knecht:2018sci, eichmann:2019bqf, roig:2019reh, colangelo:2014qya, Blum:2019ugy, Aoyama:2012wk, atoms7010028, Czarnecki:2002nt, Gnendiger:2013pva}, where the main uncertainty of the SM prediction comes from the hadronic vacuum polarization contribution.
This leads to
\beq \label{eq:Delta}
\Delta a_\mu=a_\mu^{\rm{exp}}-a_\mu^{\rm{SM}}=(251\pm 59)\times 10^{-11}\,.
\eeq
Although further scrutiny of this anomaly is needed (see e.g.~\cite{Crivellin:2020zul, Borsanyi:2020mff, Lehner:2020crt, Colangelo:2020lcg}), it is interesting to ask what this anomaly may imply for new physics~\cite{Czarnecki:2001pv, Jegerlehner:2009ry, Lindner:2016bgg}. In the SM, the contributions to the muon anomalous magnetic moment are chirally suppressed by the muon mass. Such a suppression can be lifted in the presence of new physics opening up the possibility to indirectly probe high new physics scales. Known examples include lepto-quark contributions that in some models can be enhanced by the ratio of top mass to muon mass, $m_t/m_\mu$, or contributions in the minimal supersymmetric Standard Model (MSSM) that are enhanced by $\tan\beta$, the ratio of the vacuum expectation values (vevs) of the two Higgs doublets of the MSSM. Still, the typical scale of supersymmetric (SUSY) particles required to fully address the anomaly is in the few hundred GeV range. A crucial limiting factor in the MSSM is an upper bound on $\tan\beta$ that arises from demanding perturbative Yukawa couplings of the bottom quark and the tau lepton.

In this paper, we formulate a new SUSY scenario, the flavorful supersymmetric Standard Model (FSSM). In this scenario, the contributions to $a_\mu$ can be more than an order of magnitude larger than in the MSSM. The scenario corresponds to the supersymmetrized version of the F2HDM. Its richer Higgs sector can consistently accommodate Higgsinos with $O(1)$ couplings to muons, leading to a strong enhancement of 1-loop SUSY contributions to $a_\mu$.
The paper is organized as follows: In section~\ref{sec:MSSM}, we briefly review the MSSM contributions to $a_\mu$ and show that sleptons in the few hundred GeV mass range are generically preferred. In section~\ref{sec:model}, we present our model, a SUSY version of the SM with an extended scalar sector, and discuss the features most relevant in the context of $(g-2)_\mu$. In section~\ref{sec:FMSSM} we detail the contributions to $a_\mu$ in our model and show that smuons as heavy as 6\,TeV can be responsible for the observed discrepancy. In section~\ref{sec:implications} we comment on further phenomenological implications of the model. We cover both indirect probes like lepton flavor violating tau decays and direct searches for electroweak SUSY particles at the LHC. 
Section~\ref{sec:conclusions} is reserved for conclusions and an outlook. In appendix~\ref{app:loop}, we collect the loop functions entering the several contributions to $(g-2)_\mu$.

\section{Muon Anomalous Magnetic Moment in the MSSM} \label{sec:MSSM}

We start by briefly reviewing the well known 1-loop slepton contributions to the anomalous magnetic moment of the muon in the MSSM~\cite{Chattopadhyay:1995ae,Moroi:1995yh}. We will neglect possible CP violating phases as they are strongly constrained by the non-observation of an electric dipole moment of the electron. We will also assume that the slepton soft masses are flavor conserving. Large flavor mixing between smuons and staus could in principle lead to chirally enhanced contributions to the anomalous magnetic moment of the muon~\cite{Masina:2002mv,Girrbach:2009uy}. However, taking into account the stringent constraints from $\tau \to \mu \gamma$~\cite{Aubert:2009ag,Abdesselam:2021cpu} it is found that flavor changing effects are negligibly small~\cite{Giudice:2012ms}. In presenting the MSSM contributions, it is convenient to distinguish loops with binos and winos. For masses of supersymmetric particles $m_\text{SUSY}^2 \gg m_Z^2$, one finds to a very good approximation that the two contributions are given by
\begin{eqnarray}
&& \Delta a_\mu^\text{MSSM} = \Delta a_\mu^{\tilde b} + \Delta a_\mu^{\tilde w}~, \\
&& \Delta a_\mu^{\tilde b} = \frac{g^{\prime\,2}}{192\pi^2} \frac{m_\mu^2}{m_{\tilde \mu_L}^2} \frac{M_1 \mu}{m_{\tilde \mu_L}^2}  \frac{t_\beta}{1 + \epsilon_\ell t_\beta} \left( 2 f_1(x_1,x_R) + f_2(x_1,x_\mu) -\frac{2}{x_R^2} f_2(y_1,y_\mu) \right)~, \\
&&  \Delta a_\mu^{\tilde w} = \frac{5g^2}{192\pi^2} \frac{m_\mu^2}{m_{\tilde \mu_L}^2} \frac{M_2 \mu}{m_{\tilde \mu_L}^2}  \frac{t_\beta}{1 + \epsilon_\ell t_\beta} f_3(x_2,x_\mu)~,
\end{eqnarray}
where, in the last equation, we have used the $SU(2)_L$ condition on the muon sneutrino mass $m_{\tilde\nu_\mu}=m_{\tilde \mu_L}$.
The mass ratios are given by $x_1 = M_1^2/m_{\tilde \mu_L}^2$, $y_1 = M_1^2/m_{\tilde \mu_R}^2$, $x_\mu = \mu^2/m_{\tilde \mu_L}^2$, $y_\mu = \mu^2/m_{\tilde \mu_R}^2$, $x_2 = M_2^2/m_{\tilde \mu_L}^2$, and $x_R = m_{\tilde \mu_R}^2/m_{\tilde \mu_L}^2$. In the above expressions $g$ and $g^\prime$ denote the $SU(2)_L$ and $U(1)_Y$ gauge couplings, $M_2$ and $M_1$ are the corresponding gaugino masses, $\mu$ is the Higgsino mass, and $m_{\tilde \mu_L}$, $m_{\tilde \mu_R}$ are the soft masses of the second generation slepton doublet and singlet, respectively. The parameter $\tan\beta = t_\beta = v_u/v_d$ is the ratio of the two Higgs vevs. 
We normalize the loop functions such that $f_1(1,1) = f_2(1,1) = f_3(1,1) = 1$. For completeness, their explicit expressions are given in the appendix~\ref{app:loop}. 
The parameter $\epsilon_\ell$ arises from $\tan\beta$-enhanced threshold corrections to the muon mass. It is given by \cite{Marchetti:2008hw,Bach:2015doa} (see also \cite{Hall:1993gn,Hempfling:1993kv,Carena:1994bv,Pierce:1996zz})
\begin{equation}
 \epsilon_\ell = \frac{g^{\prime\,2}}{64\pi^2} \frac{M_1 \mu}{m_{\tilde \mu_L}^2} \left(2g(x_1,x_R) + g(x_1,x_\mu) - \frac{2}{x_R} g(y_1,y_\mu)  \right) -\frac{3g^2}{64\pi^2} \frac{M_2 \mu}{m_{\tilde \mu_L}^2} g(x_2,x_\mu) ~,
\end{equation}
where the loop function is given in the appendix~\ref{app:loop} and it is normalized such that $g(1,1)=1$. For a generic point in MSSM parameter space $\epsilon_\ell \sim 10^{-3}$, and its effect becomes relevant only for very large $\tan\beta$. 

The dominant contribution to $(g-2)_\mu$ typically comes from the wino loops. In the limit that all SUSY masses are equal and neglecting the threshold corrections, the above expressions give
\begin{equation}
 \Delta a_\mu^{\rm MSSM} \simeq 260 \times 10^{-11} \times \left(\frac{t_\beta}{50} \right) \times \left(\frac{500\,{\rm GeV}}{m_{\rm SUSY}} \right)^2 ~.
\end{equation}
As it is evident from the above equation, even for large values of $\tan\beta \simeq 50$, the typical mass scale of the involved supersymmetric particles (sleptons and electroweakinos) is below 1\,TeV. This is confirmed by our numerical results in Figure \ref{fig:plot} (see the blue and purple shaded regions). The fact that an explanation of $(g-2)_\mu$ prefers a light spectrum of sleptons and electroweakinos has been re-emphasized recently in several studies of the MSSM~\cite{Aloni:2021wzk, Endo:2021zal, Iwamoto:2021aaf, Gu:2021mjd, VanBeekveld:2021tgn, Yin:2021mls, Wang:2021bcx, Chakraborti:2021dli, Cox:2021gqq, Han:2021ify, Baum:2021qzx, Athron:2021iuf, Aboubrahim:2021rwz, Chakraborti:2021bmv, baer2021anomalous} and of MSSM extensions~\cite{Abdughani:2021pdc, Heinemeyer:2021zpc}.
It is possible to accommodate the preferred value for $\Delta a_\mu$ for a somewhat heavier spectrum ($m_{{\rm{SUSY}}}\gtrsim 1$ TeV) in corners of parameter space with either a very large $\mu$ term~\cite{Endo:2021zal, Gu:2021mjd, Athron:2021iuf} or with very large values of $\tan\beta$. However, for large values of $\mu$, the MSSM scalar potential can develop charge breaking minima and vacuum stability considerations strongly constrain the parameter space. For very large values of $\tan\beta \gtrsim 70$ the bottom and tau Yukawa couplings develop Landau poles before the GUT scale, see e.g.~\cite{Altmannshofer:2010zt}.  

In the following, we present a supersymmetric extension of the Standard Model that can accommodate the measured $(g-2)_\mu$ with multi-TeV sleptons. The corresponding region of parameter space is completely safe from vacuum stability constraints and all Yukawa couplings remain perturbative.

\section{The MSSM with a Flavorful Higgs Sector}\label{sec:model}

We supersymmetrize the flavorful 2HDM.
Instead of the usual two chiral superfields $\hat H_u,\hat H_d$ of the MSSM, we introduce four chiral superfields $\hat H_u, \hat H_u^\prime, \hat H_d, \hat H_d^\prime$ (see also~\cite{Escudero:2005hk, Kawase:2011az, Dutta:2018yos} for other models with more than two Higgs doublets). Under the $SU(3)_c \times SU(2)_L \times U(1)_Y$ gauge symmetry, these superfields transform as $\hat H_u, \hat H_u^\prime \sim (\mathbf{1}, \mathbf{2})_{+\frac{1}{2}}$ and $\hat H_d, \hat H_d^\prime \sim (\mathbf{1}, \mathbf{2})_{-\frac{1}{2}}$. The superpotential of the model is given by
\begin{multline}
W= \mu_1\hat H_u\hat H_d+\mu_2\hat H_u^\prime\hat H_d^\prime+\mu_3\hat H_u^\prime\hat H_d+\mu_4\hat H_u\hat H_d^\prime\\
 + (Y_u \hat H_u + Y^\prime_u \hat H_u^\prime ) \hat Q \hat U^c + ( Y_d \hat H_d + Y^\prime_d \hat H_d^\prime ) \hat Q \hat D^c + (Y_\ell \hat H_d + Y^\prime_\ell \hat H_d^\prime) \hat L \hat E^c \,.
\end{multline}
It contains four independent $\mu$-terms as well as the Yukawa couplings $Y_f$ and $Y_f^\prime$ of the Higgs doublets to the matter superfields. In the following we will denote this model as the flavorful supersymmetric Standard Model or FSSM.

We assume that the neutral components of the Higgs scalars acquire vevs given by $v_u$, $v_u^\prime$, $v_d$, and $v_d^\prime$, such that $v_u^2 + v_d^2 + v_u^{\prime\,2} + v_d^{\prime\,2} = v^2 = (246\,\text{GeV})^2$.  
In addition to the usual vev ratio $\tan\beta = t_\beta = v_u/v_d$, we also introduce the ratios $\tan\beta_u = t_{\beta_u}= v_u/v^\prime_u$ and $\tan\beta_d = t_{\beta_d} = v_d/v^\prime_d$. 
Generic Yukawa couplings $Y_f$ and $Y_f^\prime$ violate the principle of natural flavor conservation. They can lead to large neutral Higgs mediated flavor changing neutral currents and are therefore strongly constrained. In the following we will consider the ``flavorful'' ansatz for the Yukawa couplings as suggested in~\cite{Altmannshofer:2015esa} that avoids the most stringent flavor constraints due to an approximate flavor symmetry~\cite{Altmannshofer:2015esa}. In this ansatz, the doublets $\hat H_u$, $\hat H_d$ couple exclusively to third generation fermions, while $\hat H_u^\prime$, $\hat H_d^\prime$ provide masses for the first and second generations. In particular, the muon mass is proportional to the vev $v_d^\prime$ and the corresponding muon Yukawa coupling $Y_{\mu\mu}^\prime$. The muon Yukawa is thus enhanced by a factor $\tan\beta \tan\beta_d$ compared to the usual $\tan\beta$ enhancement present in the MSSM.

In the case of all three leptons, we have (neglecting SUSY threshold effects)\footnote{A similar structure can be implemented in the quark sector.}
\begin{equation} \label{eq:Yell}
Y_\ell^\prime  \simeq \frac{\sqrt{2}}{v_d^\prime}
\begin{pmatrix}
~m_e~& ~x_{e\mu} m_e ~& ~ x_{e\tau} m_e~ \\
~x_{\mu e} m_e ~& ~m_\mu ~& ~x_{\mu\tau}m_\mu~ \\
~x_{\tau e} m_e~& ~x_{\tau\mu} m_\mu~ & ~x_{\tau\tau} m_\mu~
\end{pmatrix} ~,~~~~~~~~
 Y_\ell  \simeq \frac{\sqrt{2}}{v_d}
\begin{pmatrix}
~0~& ~0 ~& ~0~ \\
~0 ~& ~0 ~& ~0~ \\
~0~&~ 0~ & ~m_\tau~
\end{pmatrix} ~. 
\end{equation}
The rank-1 Yukawa coupling $Y_\ell$ preserves an $SU(2)^2 = SU(2)_L \times SU(2)_E$ flavor symmetry acting of the first two generations of left-handed and right-handed lepton fields. The structure of the second Yukawa coupling $Y_\ell^\prime$ follows the ansatz in~\cite{Altmannshofer:2015esa}: a $2\times 2$ block with entries of the order of the muon mass with the remaining entries of the order of the electron mass. The $x_{ij}$ parameters indicate how much the Yukawa coupling $Y_{\ell}^\prime$ differs from such an ansatz. Generically, one might expect the $x_{ij}$ are of order one, but they can be much smaller as well.
Assuming that soft SUSY breaking is flavor universal, the $SU(2)^2$ flavor symmetry is minimally broken by the second Yukawa coupling $Y_\ell^\prime$, implying that flavor changing effects between the second and first generation leptons are strongly suppressed.  The parameters $x_{\tau\mu}$ and $x_{\mu\tau}$ can be constrained by the experimental bounds on flavor violating tau decays like $\tau \to \mu\gamma$ and $\tau\to3\mu$, while the corresponding parameters with electrons, $x_{\tau e}$ and $x_{e\tau}$, can be constrained from data on $\tau \to e$ transitions. Due to the $SU(2)^2$ flavor symmetry, the parameters $x_{\mu e}$ and $x_{e\mu}$ are unobservable. Effects in the highly constrained $\mu \to e$ transitions like $\mu\to e\gamma$ can be expected only if the products $x_{\mu\tau}x_{\tau e}$ or $x_{e\tau}x_{\tau \mu}$ are sizable (see section \ref{sec:implications} for more details).

In addition to the leptonic Yukawa couplings shown above, the most relevant ingredients for the discussion of $(g-2)_\mu$ are the smuon masses as well as the chargino and neutralino masses.
For the smuon mass matrix we find after electroweak symmetry breaking
\beq \label{eq:smuon_mass}
M_{\tilde\mu}^2=\left(\begin{array}{cc}
m_{\tilde \mu_L}^2  & - m_\mu t_\beta t_{\beta_d} (\mu_4  + \mu_2 / t_{\beta_u}) \\
- m_\mu t_\beta t_{\beta_d} (\mu_4  + \mu_2 / t_{\beta_u}) & m_{\tilde \mu_R}^2
\end{array}\right)\,,
\eeq
where 
we have neglected the small D-term contributions, as well as the contributions from soft trilinear terms. Neglecting the trilinear terms is a good approximation as long as $v_d^\prime \ll v_u$.
Note that in eq.~(\ref{eq:smuon_mass}) we neglected left-right mixing between smuons and staus proportional to $x_{\tau\mu}$ and $x_{\mu\tau}$. Such mixing is of no relevance to the calculation of $(g-2)_\mu$. We will comment on its effect on $\tau \to \mu\gamma$ in section~\ref{sec:implications}.

The model features an extended electroweakino sector, because of the additional Higgsinos. It contains three charginos and six neutralinos with the following mass matrices
\beq
M_{\chi^\pm}=\left(\begin{array}{ccc}
M_2 & \frac{g}{\sqrt 2} v_u & \frac{g}{\sqrt 2} v_u^\prime\\
\frac{g}{\sqrt 2} v_d & \mu_1 & \mu_3\\
\frac{g}{\sqrt 2} v_d^\prime & \mu_4 & \mu_2
\end{array}\right)\,,~~
M_{\chi^0}=\left(\begin{array}{cccccc}
M_1 & 0 & -\frac{g^\prime}{2}v_d & \frac{g^\prime}{2}v_u & -\frac{g^\prime}{2}v_d^\prime & \frac{g^\prime}{2}v_u^\prime\\
0 & M_2 & \frac{g}{2}v_d & -\frac{g}{2}v_u & \frac{g}{2}v_d^\prime & -\frac{g}{2}v_u^\prime\\
-\frac{g^\prime}{2}v_d & \frac{g}{2}v_d & 0 & -\mu_1 & 0 & -\mu_3\\
\frac{g^\prime}{2}v_u &-\frac{g}{2}v_u & -\mu_1 & 0 & -\mu_4 & 0\\
-\frac{g^\prime}{2}v_d^\prime&  \frac{g}{2}v_d^\prime & 0 &-\mu_4 & 0 & -\mu_2\\
 \frac{g^\prime}{2}v_u^\prime& -\frac{g}{2}v_u^\prime & -\mu_3 & 0 & -\mu_2 & 0
\end{array}\right)\,,
\eeq
where we have considered the basis $(\tilde W^\pm,\tilde H^\pm,\tilde H^{\prime\pm})$ and $(\tilde B, \tilde W^0, \tilde H_d^0, \tilde H_u^0, \tilde H_d^{\prime0}, \tilde H_u^{\prime0})$ for charginos and neutralinos, respectively. 
For the calculation of $(g-2)_\mu$ it is convenient to rotate the Higgsino states to diagonalize the $2\times 2$ Higgsino sub-matrix:
\beq
\left(\begin{array}{cc}
c_d & s_d\\
-s_d & c_d
\end{array}\right)\left(\begin{array}{cc}
\mu_1 & \mu_3\\
\mu_4 & \mu_2
\end{array}\right)
\left(\begin{array}{cc}
c_u & s_u\\
-s_u & c_u
\end{array}\right)=\left(
\begin{array}{cc}
\mu & 0\\
0 & \tilde\mu
\end{array}\right)\,,
\eeq
where we have introduced the mixing angles $c_{d,u}\equiv \cos\theta_{d,u}$, $s_{d,u}\equiv \sin\theta_{d,u}$. 
For gaugino and Higgsino masses sufficiently above the electroweak scale, the masses of the three charginos and of the six neutralinos are approximately $m_{\chi^\pm_i} \simeq (M_2,\mu,\tilde \mu)$ and $m_{\chi^0_i} \simeq (M_1,M_2,\mu,\mu,\tilde \mu,\tilde \mu)$.
In the following section we will report the contributions to the anomalous magnetic moment of the muon in terms of the Higgsino parameters $\theta_u,\theta_d$ and $\mu, \tilde \mu$.

\section{\boldmath FSSM Contributions to \texorpdfstring{$(g-2)_\mu$}{(g-2)mu}} \label{sec:FMSSM}

\begin{figure}[tb]
\centering
\includegraphics[width=0.7 \textwidth]{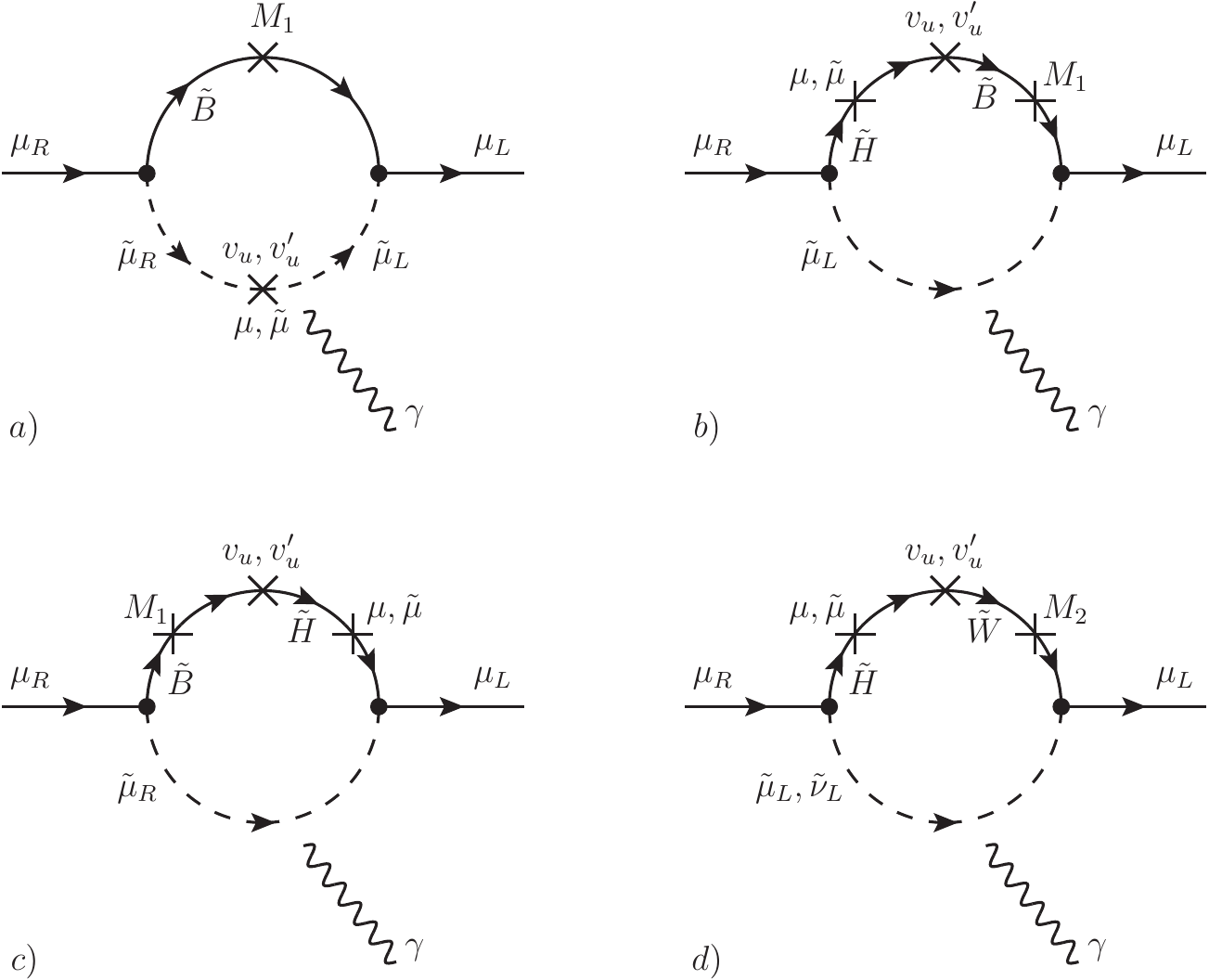}
\caption{The leading 1-loop contributions to the anomalous magnetic moment of the muon. The external photon has to be attached in all possible ways to the loops. Diagrams a), b), and c) involve binos, while diagram d) involves winos. The threshold corrections to the muon mass correspond to analogous diagrams without the external photon. 
\label{fig:diagrams}}
\end{figure}

Similar to the MSSM, it is possible to split the contributions to the anomalous magnetic moment of the muon into bino mediated and wino mediated contributions. For supersymmetric particles that are sufficiently heavier than the electroweak scale, $m_\text{SUSY}^2 \gg m_Z^2$, we find 
\begin{eqnarray}\label{eq:gminus2bino}
&&\Delta a_\mu^\text{FSSM} = \Delta a_\mu^{\tilde b} + \Delta a_\mu^{\tilde w} ~, \\
&& \Delta a_\mu^{\tilde b} = \frac{g^{\prime\,2}}{192\pi^2} \frac{m_\mu^2}{m_{\tilde \mu_L}^2}   \frac{t_\beta t_{\beta_d}}{1 + \epsilon_\ell t_\beta t_{\beta_d}} \\ \nonumber
&&~~~~~~~~~~~ \times \left[ s_d\left(c_u - \frac{s_u}{t_{\beta_u}}\right) \frac{M_1 \mu}{m_{\tilde \mu_L}^2} \left( 2 f_1(x_1,x_R) + f_2(x_1,x_\mu) -\frac{2}{x_R^2} f_2(y_1,y_\mu) \right) \right.  \\\nonumber
&&~~~~~~~~~~   \left. + c_d\left(s_u + \frac{c_u}{t_{\beta_u}}\right) \frac{M_1 \tilde \mu}{m_{\tilde \mu_L}^2} \left( 2 f_1(x_1,x_R) + f_2(x_1,x_{\tilde \mu}) -\frac{2}{x_R^2} f_2(y_1,y_{\tilde \mu}) \right) \right] ~,\\\label{eq:gminus2wino}
 && \Delta a_\mu^{\tilde w} = \frac{5g^2}{192\pi^2} \frac{m_\mu^2}{m_{\tilde \mu_L}^2} \frac{t_\beta t_{\beta_d}}{1 + \epsilon_\ell t_\beta t_{\beta_d}} \left[s_d\left(c_u - \frac{s_u}{t_{\beta_u}}\right) \frac{M_2 \mu}{m_{\tilde \mu_L}^2} f_3(x_2,x_{\mu}) \right.  \\\nonumber
 &&~~~~~~~~~~~~~~~~~~~~~~~~~~~~~~~~~~~~~~~~~~~~~~~~~~~~~~~~~~   \left. + c_d\left(s_u + \frac{c_u}{t_{\beta_u}}\right) \frac{M_2 \tilde \mu}{m_{\tilde \mu_L}^2} f_3(x_2,x_{\tilde \mu}) \right] ~,
\end{eqnarray}
with the threshold correction parameter $\epsilon_\ell$ given by
\begin{multline}
 \epsilon_\ell = \frac{g^{\prime\,2}}{64\pi^2} \left[ s_d\left(c_u - \frac{s_u}{t_{\beta_u}}\right) \frac{M_1 \mu}{m_{\tilde \mu_L}^2} \left(2g(x_1,x_R) + g(x_1,x_\mu) - \frac{2}{x_R} g(y_1,y_\mu)  \right) \right.  \\
 \left. + c_d\left(s_u + \frac{c_u}{t_{\beta_u}}\right) \frac{M_1 \tilde \mu}{m_{\tilde \mu_L}^2} \left(2g(x_1,x_R) + g(x_1,x_{\tilde \mu}) - \frac{2}{x_R} g(y_1,y_{\tilde \mu})  \right) \right]  \\
  -\frac{3g^2}{64\pi^2} \left[s_d\left(c_u - \frac{s_u}{t_{\beta_u}}\right) \frac{M_2 \mu}{m_{\tilde \mu_L}^2} g(x_2,x_\mu) + c_d\left(s_u + \frac{c_u}{t_{\beta_u}}\right) \frac{M_2 \tilde \mu}{m_{\tilde \mu_L}^2} g(x_2,x_{\tilde \mu}) \right] ~,
\end{multline}
where, similarly to $x_\mu$ and $y_\mu$, we have defined $x_{\tilde\mu}=\tilde\mu^2/m_{\tilde\mu_L}^2$ and $y_{\tilde\mu}=\tilde\mu^2/m_{\tilde\mu_R}^2$. Note that the loop functions are identical to the MSSM case. In fact, the expressions above largely resemble the MSSM result shown in section~\ref{sec:MSSM}. The relevant Feynman diagrams containing Higgsinos, binos, winos, and sleptons are shown in Figure~\ref{fig:diagrams}\footnote{In addition, the model also predicts contributions from 1-loop diagrams containing leptons and Higgs bosons. However, such contributions to $\Delta a_\mu$ are not chirally enhanced by $\tan\beta$ factors and therefore can be neglected.}. As in the MSSM, the dominant contribution typically comes from the loops containing a wino. If all SUSY masses are set equal, the wino loops dominate over the bino loops by a factor $5 g^2/g^{\prime\,2} \simeq 17$. In contrast to the MSSM, we find two sets of contributions that are proportional to either one of the Higgsino mass eigenvalues $\mu$ and $\tilde \mu$. The main difference to the MSSM is the overall proportionality to the product of the vev ratios $\tan\beta \times \tan\beta_d = v_u/v_d^\prime$. The additional factor $\tan\beta_d$ can increase the contributions in our setup by an order of magnitude or more compared to the MSSM. Note that the threshold correction remains of order $\epsilon_\ell \sim 10^{-3}$ in our setup. However, as it is multiplied by the product $\tan\beta \times \tan\beta_d$ (see Eqs. (\ref{eq:gminus2bino}), (\ref{eq:gminus2wino})), it can have an $O(1)$ impact on the contribution to the $(g-2)_\mu$ (we show its effect explicitly in Figure \ref{fig:plot} below). Setting all SUSY masses equal and assuming maximal mixing in the Higgsino sector $s_d = c_d = s_u = c_u = 1/\sqrt{2}$, the threshold correction corresponds to the following 1-loop correction to the muon mass
\begin{eqnarray} \label{eq:delta_mmu_approx}
 \frac{\Delta m_\mu^\text{1-loop}}{m_\mu^\text{tree}} = \epsilon_\ell t_\beta t_{\beta_d} \simeq - 0.54 \times \left( \frac{t_\beta}{20} \right) \times \left( \frac{t_{\beta_d}}{15} \right) ~.
\end{eqnarray}
Note that for this benchmark case, the correction is identical to the correction in the MSSM (see e.g.~\cite{Bach:2015doa}) up to the additional factor $\tan\beta_d$. Away from the limit $s_d = c_d = s_u = c_u = 1/\sqrt{2}$, the correction to the muon mass is modified by the factor $(s_d(c_u-s_u/t_{\beta_u}) +c_d(s_u +c_u/t_{\beta_u} )) \sim \mathcal O(1)$.

A relevant bound on the size of $\tan\beta_d$ is given by perturbativity considerations. Yukawa couplings that are larger than $O(1)$ at the TeV scale develop Landau poles before reaching the GUT scale. Requiring that the muon Yukawa of the Higgs field $H_d^\prime$ stay perturbative up to the GUT scale, leads to the approximate bound\footnote{We estimate that the perturbativity bound on the muon Yukawa in our model will be similar to the bound on the tau Yukawa in the MSSM. In Eq. (\ref{eq:perturbativitybound}), we therefore quote the bound on the tau Yukawa coupling at a scale of 1\,TeV that has been found in the MSSM requiring the Yukawa coupling to be smaller than $\sqrt{4\pi}$ at the GUT scale~\cite{Altmannshofer:2010zt}. A dedicated renormalization group study of the full set of third and second generation Yukawa couplings and of the gauge couplings would be required to establish a precise bound on the muon Yukawa coupling in our model, but we do not expect the result to change significantly.} 
\begin{equation}\label{eq:perturbativitybound}
 Y^\prime_{\mu\mu} \simeq \frac{\sqrt{2} m_\mu}{v} \frac{t_\beta t_{\beta_d}}{1 + \epsilon_\ell t_\beta t_{\beta_d}} \lesssim 0.7 ~.
\end{equation}
Differently from the MSSM, the requirement of perturbativity of the tau and bottom Yukawa couplings sets weaker bounds on the values of $\tan\beta$ and $\tan\beta_d$.
In the limit that all SUSY masses are equal and assuming maximal mixing in the Higgsino sector $s_d = c_d = s_u = c_u = 1/\sqrt{2}$, we find
\begin{eqnarray} \label{eq:delta_amu_approx_1}
 \Delta a_\mu &\simeq& 220 \times 10^{-11} \times \left(\frac{t_\beta}{20} \right)\times \left(\frac{t_{\beta_d}}{15} \right)\times \left(\frac{0.46}{1 + \epsilon_\ell t_{\beta} t_{\beta_d}} \right) \times \left(\frac{2.0\,{\rm TeV}}{m_{\rm SUSY}} \right)^2 \\  \label{eq:delta_amu_approx_2}
 &\simeq& 240 \times 10^{-11} \times \left(\frac{Y^\prime_{\mu\mu}}{0.7} \right) \times \left(\frac{2.5\,{\rm TeV}}{m_{\rm SUSY}} \right)^2 ~.
\end{eqnarray}
Analogously to the corrections to the muon mass in~(\ref{eq:delta_mmu_approx}), for generic $O(1)$ mixing of the FSSM Higgsinos, the results in~(\ref{eq:delta_amu_approx_1}) and~(\ref{eq:delta_amu_approx_2}) have to be multiplied by the factor $s_d(c_u-s_u/t_{\beta_u}) +c_d(s_u +c_u/t_{\beta_u})$.
In the MSSM, similar equations for $\Delta a_\mu$ hold up to the additional factor $\tan\beta_d$. Comparably large contributions to $\Delta a_\mu$ are therefore in principle possible in the MSSM in the ultra-large $\tan\beta$ limit (see e.g.~\cite{Bach:2015doa}). However, as mentioned in section~\ref{sec:MSSM}, the required values of $\tan\beta$ for multi-TeV SUSY particles are very strongly constrained by perturbativity considerations~\cite{Altmannshofer:2010zt}. Those constraints are relaxed in the FSSM, and very large values for the product $\tan\beta\tan\beta_d$ are viable.

Keeping in mind the estimated bound on the muon Yukawa discussed above, we find that the generic scale of supersymmetric particles can be larger by a factor of $\simeq 5$ compared to the MSSM, while still producing the desired effect in the anomalous magnetic moment of the muon.

\begin{figure}[tb]
\centering
\includegraphics[width=0.6 \textwidth]{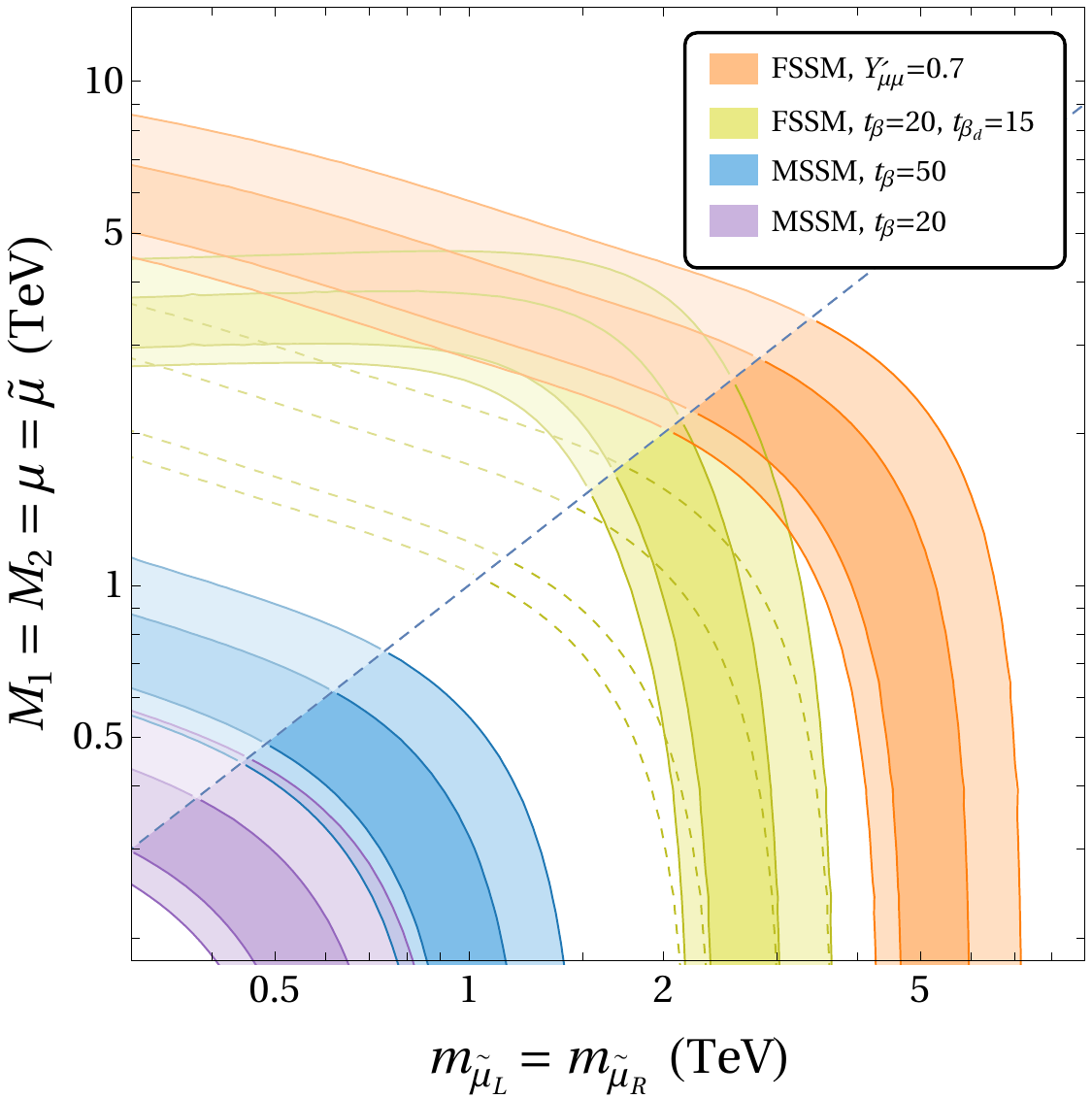}
\caption{Regions of SUSY parameter space that are preferred by the anomalous magnetic moment of the muon at the $1\sigma$ and $2\sigma$ level in several benchmark models. Above the dashed diagonal line, the smuons are lighter than any of the gauginos and Higgsinos. In blue and purple, we present two MSSM scenarios; in yellow and orange two FSSM scenarios. The dashed yellow lines show the corresponding FSSM scenario neglecting the SUSY threshold corrections to the muon mass.
\label{fig:plot}}
\end{figure}

In Figure~\ref{fig:plot}, we show the regions of SUSY masses that are preferred by the anomalous magnetic moment of the muon in several benchmark cases. For simplicity, we assume that the masses of the left-handed and right-handed smuons are equal, as do we for the masses of the bino, the wino, and the Higgsinos. In the colored bands we find agreement with (\ref{eq:Delta}) at the $1\sigma$ and $2\sigma$ level. The purple and blue bands correspond to the MSSM with $\tan\beta = 20$ and $\tan\beta = 50$, respectively. We see that for $\tan\beta =50$, sleptons can be at most at around $1$\,TeV, if gauginos and Higgsinos are in the few hundred GeV range. The yellow and orange bands show two benchmark scenarios in the FSSM assuming a generic $O(1)$ mixing in the Higgsino sector $s_d \sim c_d \sim s_u \sim c_u \sim 1/\sqrt{2}$. The yellow band corresponds to moderate values for $\tan\beta = 20$ and $\tan\beta_d = 15$. In such a scenario the smuons can be as heavy as $3$\,TeV while still explaining $(g-2)_\mu$. The dashed yellow lines show the region favored by $(g-2)_\mu$ in this FSSM scenario neglecting the SUSY threshold corrections to the muon mass (i.e. $\epsilon_\ell$ is set to zero). We clearly see that the threshold corrections have an order 1 impact and cannot be neglected.\footnote{In this context, it might be interesting to consider a scenario in which the entire muon mass is radiatively generated. In such a case one expects that the $(g-2)_\mu$ anomaly can be explained with new physics masses at around 2 TeV. The radiative muon mass scenario has been analyzed in the MSSM~\cite{Bach:2015doa,Crivellin:2010ty} and it is strongly constrained by perturbativity arguments.} Finally, the orange band shows a scenario in which we choose a large muon Yukawa coupling, $Y_{\mu\mu}^\prime = 0.7$, that we estimate to be close to the bound from demanding perturbativity up to the GUT scale. In this case, smuons can be as heavy as $6$\,TeV. 

\section{Phenomenological Implications} \label{sec:implications}

In addition to the contributions to $\Delta a_\mu$, the FSSM predicts contributions to the anomalous magnetic moments of the electron and of the tau, $\Delta a_e$ and $\Delta a_\tau$. Due to the minimally broken $SU(2)^2$ lepton symmetry, the setup predicts the relation
\begin{equation}
 \Delta a_e \simeq \frac{m_e^2}{m_\mu^2} \Delta a_\mu \simeq 5.8 \times 10^{-14} \times \left( \frac{\Delta a_\mu}{251 \times 10^{-11}} \right) ~,
\end{equation}
which is almost an order of magnitude smaller than the uncertainties of the experimental determination~\cite{Hanneke:2008tm}, as well as the uncertainties of the SM prediction that depends crucially on the value of the fine structure constant~\cite{Parker:2018vye, Morel:2020dww}.
There is no strict correlation of $\Delta a_\mu$ and $\Delta a_\tau$, but generically we expect
\begin{equation}
 \Delta a_\tau \sim \frac{m_\tau^2}{m_\mu^2} \frac{1}{t_{\beta_d}} \Delta a_\mu \simeq 4.7 \times 10^{-8} \times \left( \frac{15}{t_{\beta_d}} \right) \times \left( \frac{\Delta a_\mu}{251 \times 10^{-11}} \right) ~,
\end{equation}
which is far below foreseeable experimental sensitivities \cite{Abdallah:2003xd}.

Similar to the 1-loop slepton contributions to anomalous magnetic moments, the FSSM setup also gives contributions to the radiative lepton decays $\tau \to \mu \gamma$, $\tau \to e \gamma$, and $\mu\to e \gamma$. While the relevant off-diagonal couplings $x_{\tau e}$, $x_{e\tau}$, $x_{\tau \mu}$, and $x_{\mu\tau}$ in the lepton Yukawa matrix~(\ref{eq:Yell}) do not enter the predictions for $\Delta a_\mu$ at the considered level of accuracy, it is nonetheless interesting to explore their implications.
In the limit in which both left-handed and right-handed smuons and staus have equal masses we find simple relations between the 1-loop slepton contributions to the anomalous magnetic moment of the muon and the branching ratios of the decays $\tau \to \mu \gamma$, $\tau \to e \gamma$, and $\mu\to e \gamma$
\begin{eqnarray}
 \text{BR}(\tau \to \mu\gamma) &\simeq& 24\pi^3 \alpha_\text{em} \frac{v^4}{m_\mu^4} \left(\Delta a_\mu \right)^2 \big(x_{\tau\mu}^2 + x_{\mu\tau}^2 \big) \times \text{BR}(\tau \to \mu \nu_\tau \bar\nu_\mu) \nonumber \\
 &\simeq& 1.7 \times 10^{-8} \times \left(\frac{\Delta a_\mu}{251 \times 10^{-11}}\right)^2 \left[ \left(\frac{x_{\tau\mu}}{0.01}\right)^2 + \left(\frac{x_{\mu\tau}}{0.01}\right)^2 \right]~, \\
  \text{BR}(\tau \to e\gamma) &\simeq& 24\pi^3 \alpha_\text{em} \frac{v^4}{m_\mu^4} \frac{m_e^2}{m_\mu^2} \left(\Delta a_\mu \right)^2 \big(x_{\tau e}^2 + x_{\mu e}^2 \big) \times \text{BR}(\tau \to e \nu_\tau \bar\nu_\mu) \nonumber \\
 &\simeq& 4.1 \times 10^{-9} \times \left(\frac{\Delta a_\mu}{251 \times 10^{-11}}\right)^2 \left[ \left(\frac{x_{\tau e}}{1.0}\right)^2 + \left(\frac{x_{e\tau}}{1.0}\right)^2 \right]~, \\
 \text{BR}(\mu \to e\gamma) &\simeq& 24\pi^3 \alpha_\text{em} \frac{v^4}{m_\mu^4} \frac{m_e^2}{m_\tau^2} \left(\Delta a_\mu \right)^2 \big(x_{e\tau}^2 x_{\tau\mu}^2 + x_{\mu\tau}^2x_{\tau e}^2 \big) \nonumber \\
 &\simeq& 8.2 \times 10^{-15} \times \left(\frac{\Delta a_\mu}{251 \times 10^{-11}}\right)^2 \left[ \left(\frac{x_{e\tau} x_{\tau\mu}}{0.01}\right)^2 + \left(\frac{x_{\mu\tau}x_{\tau e}}{0.01}\right)^2 \right]~,
\end{eqnarray}
where we used BR$(\tau \to \mu \nu_\tau \bar\nu_\mu)\simeq 17.4\%$ and BR$(\tau \to e \nu_\tau \bar\nu_\mu)\simeq 17.8\%$~\cite{Zyla:2020zbs}. 

We can compare these predictions to the current experimental bounds. From the bound
BR$(\tau \to \mu \gamma)_\text{exp} < 4.2\times 10^{-8}$~\cite{Aubert:2009ag,Abdesselam:2021cpu} we see that the couplings $x_{\tau \mu}$ and $x_{\mu\tau}$ have to be of order $10^{-2}$ in order not to violate the bound from $\tau \to \mu \gamma$. The smallness of these couplings suggests that an additional lepton flavor symmetry gives structure to the lepton Yukawa coupling $Y^\prime_\ell$. If the coupling $x_{\tau e}$ or $x_{e\tau}$ is of order~$1$, the branching ratio of $\tau \to e \gamma$ is predicted below the current bound, BR$(\tau \to e \gamma)_\text{exp} < 3.3\times 10^{-8}$~\cite{Aubert:2009ag,Abdesselam:2021cpu}, but in reach of the Belle II experiment~\cite{Kou:2018nap}. Once the bound from BR$(\tau \to \mu \gamma)$ is taken into account, the $\mu \to e \gamma$ branching ratio is predicted well below the current constraint BR$(\mu \to e \gamma)_\text{exp} < 4.2\times 10^{-13}$~\cite{TheMEG:2016wtm}.

In the FSSM, we also find tree level neutral Higgs contributions to lepton flavor violating decays $\tau \to 3 \mu$, $\tau \to 3e$, and $\mu \to 3e$. The most constraining decay is expected to be $\tau \to 3 \mu$ as it involves the largest Yukawa couplings enhanced by the product $t_\beta t_{\beta_d}$. Neglecting mixing among the Higgs bosons and using the results from~\cite{Paradisi:2005tk}, we find 
\begin{eqnarray}\nonumber
 \text{BR}(\tau \to 3\mu) &\simeq& \frac{m_\mu^4}{4 m_{H_d^\prime}^4} \frac{t_\beta^4 t_{\beta_d}^4}{(1 + \epsilon_\ell t_\beta t_{\beta_d})^4} \big(x_{\tau\mu}^2 + x_{\mu\tau}^2 \big) \times \text{BR}(\tau \to \mu \nu_\tau \bar\nu_\mu) \\
 &\simeq& 1.0\times 10^{-9} \times \left(\frac{Y^\prime_{\mu\mu}}{0.7} \right)^4 \times \left(\frac{1.0\,{\rm TeV}}{m_{H_d^\prime}} \right)^4 \times \left[ \left(\frac{x_{\tau\mu}}{0.01}\right)^2 + \left(\frac{x_{\mu\tau}}{0.01}\right)^2 \right] ~.
\end{eqnarray}
Comparing to the experimental bound obtained by the Belle collaboration, BR$(\tau \to 3 \mu)_\text{exp} < 2.1\times 10^{-8}$~\cite{Hayasaka:2010np} (see also~\cite{Lees:2010ez, Aaij:2014azz}), we see that TeV scale Higgs bosons are viable and could lead to branching ratios that are accessible at Belle II and LHCb \cite{Kou:2018nap,Bediaga:2012py}.

Finally, we comment on the prospects of testing the SUSY parameter space favored by $(g-2)_\mu$ at the LHC. The sleptons necessary for addressing the $(g-2)_\mu$ anomaly are being searched for by the ATLAS and CMS collaborations. Particularly, the most relevant slepton signature is the slepton pair production $pp\to\tilde\ell\tilde\ell$, followed by the decay into a lepton and the lightest neutralino $\tilde\ell\tilde\ell\to(\ell\tilde\chi^0)(\ell\tilde\chi^0)$. The most stringent bounds on the slepton parameter space come from the analyses \cite{Sirunyan:2020eab,Aad:2019vnb} performed with the full Run II data set. These searches show that, in the case of degenerate left-handed and right-handed smuons, slepton masses as large as $\sim 600$ GeV are now generically probed if the mass splitting with the lightest neutralino $m_{\tilde\ell}-m_{\tilde\chi^0}$ is sufficiently large. In the case of a compressed spectrum, the limits are weaker and allow sleptons as light as $\sim 250$ GeV for mass splitting $m_{\tilde\ell}-m_{\tilde\chi^0}\lesssim 30$ GeV \cite{Aad:2019qnd}. 

The existing searches probe already part of the MSSM parameter space that is able to explain the $(g-2)_\mu$ anomaly. However, depending on the specific electroweakino spectrum, sizable regions of parameter space are left unexplored. For example, for the specific spectrum fixed in Figure \ref{fig:plot} ($M_1=M_2=\mu$), the aforementioned LHC slepton searches can only set a weak bound because of the dilution of the branching ratio of the slepton into the lightest neutralino and a lepton. For that spectrum, additional slepton decay modes arise including $\tilde\mu\to\mu\tilde\chi_2^0,\nu\tilde\chi_{1,2}^\pm$ with the subsequent decay of $\tilde\chi_2^0$ and $\tilde\chi_{1,2}^\pm$ to the lightest neutralino and jets or leptons. It will be interesting to search for these new slepton cascade decays in the coming years at the LHC to probe further regions of MSSM parameter space that can address the $(g-2)_\mu$ anomaly.

The slepton phenomenology in the FSSM is even richer. Due to the extended electroweakino sector, several cascade decays are possible, giving rise to signatures with multi-leptons (or jets) and missing energy. We leave the study of such signatures for future work. Due to the generically heavier slepton masses in the FSSM to address the $(g-2)_\mu$ anomaly, the sleptons that we considered in this paper are outside the reach of the LHC. We expect some of the scenarios favored by $(g-2)_\mu$ to be probed at a 100 TeV collider through a $pp\to\tilde\ell\tilde\ell\to(\ell\tilde\chi^0)(\ell\tilde\chi^0)$ search \cite{Baker:2018uox}.

\section{Conclusions and Outlook} \label{sec:conclusions}

The recent result of the Fermilab Muon g-2 collaboration confirms the longstanding discrepancy in the anomalous magnetic moment of the muon, $(g-2)_\mu$. Motivated by this result, we constructed a supersymmetric extension of the Standard Model that can give more than an order of magnitude larger contributions to the $(g-2)_\mu$ than the MSSM. The model, that we dub flavorful supersymmetric Standard Model (FSSM), is based on an extended electroweak breaking sector with Yukawa couplings that follow the flavorful ansatz suggested in~\cite{Altmannshofer:2015esa}. 

One set of Higgs doublets couples exclusively to the third generation through a rank-1 Yukawa coupling, while a second set of Higgs doublets couples also to the first and second generations. One of the Higgs fields can have an $O(1)$ coupling to muons. Loop contributions to $(g-2)_\mu$ that contain the corresponding Higgsino state are strongly enhanced. In contrast to the MSSM, where the SUSY particles generically have to be below the 1\,TeV scale, in the FSSM, the $(g-2)_\mu$ result can be comfortably explained by sleptons in the multi-TeV mass range.

We explored further phenomenological implications of the model, in particular for the anomalous magnetic moments of the electron and the tau, as well as for lepton flavor violation. We found that existing bounds on $\tau\to\mu\gamma$ already give relevant constraints on the lepton flavor violating Yukawa couplings of the model. The $\tau \to 3\mu$ and $\tau\to e \gamma$ decays might be in reach of running experiments. 

While LHC searches for sleptons already constrain some of the MSSM explanations of the $(g-2)_\mu$ discrepancy, directly probing the multi-TeV sleptons of the FSSM will require a future higher energy collider.

There are several possible future directions to further explore the FSSM. The rank-1 ansatz for the Yukawa couplings has been successfully implemented in the quark sector in non-supersymmetric versions of the model~\cite{Altmannshofer:2015esa,Altmannshofer:2016zrn,Altmannshofer:2018bch,Altmannshofer:2019ogm,Altmannshofer:2017uvs}. It has been shown that in the considered scenarios quark flavor changing neutral currents can be relatively easily in agreement with experimental bounds. In a supersymmetric version, we expect interesting Higgsino mediated effects in chirality suppressed processes like $b \to s\gamma$ and $B_s \to \mu^+ \mu^-$ and possibly even for the lepton flavor universality ratios $R_K$ and $R_{K^*}$. A study of those effects will be presented elsewhere.

If the rank-1 ansatz for the Yukawa couplings is implemented both in the lepton and quark sectors, a scenario with $\tan\beta \sim 50$, $\tan\beta_u \sim 100$, and $\tan\beta_d \sim 10$ can address all hierarchies among the third and second generation quark and lepton masses without any pronounced hierarchy in Yukawa couplings. These are the values of $\tan\beta$ and $\tan\beta_{d}$ that allow multi-TeV sleptons to address the $(g-2)_\mu$ anomaly. In such a region of parameter space, third and second generation Yukawa unification might be possible at the GUT scale. For the first generation, one could entertain the possibility of radiative mass generation to explain the smallness of the up quark, down quark, and electron mass.
We leave these studies to future work.

\section*{Acknowledgements}
We thank Kaustubh Agashe for discussions. The research of W. A. is supported by the U.S. Department of Energy grant number DE-SC0010107. The research of SG is supported in part by the National Science Foundation CAREER grant PHY- 1915852 and by the National Science Foundation under Grant No. NSF PHY-1748958.

\appendix
\section{Loop Functions} \label{app:loop}

In this appendix, we collect the loop functions that enter the expressions for the SUSY contributions to $(g-2)_\mu$ discussed in sections~\ref{sec:MSSM} and~\ref{sec:FMSSM}

\begin{eqnarray}\nonumber
f_1(x,y) &=& \frac{6(y-3x^2+x(1+y))}{(1-x)^2(x-y)^2} - \frac{12x(x^3+y-3xy+y^2)\log x}{(1-x)^3(x-y)^3} + \frac{12xy\log y}{(1-y)(x-y)^3} ~, \\
f_2(x,y) &=& \frac{6(x+y+xy-3)}{(1-x)^2(1-y)^2} -\frac{12x\log x}{(1-x)^3(x-y)} + \frac{12y \log y}{(1-y)^3(x-y)} ~, \\\nonumber
f_3(x,y) &=& \frac{6(13-7(x+y)+xy)}{5(1-x)^2(1-y)^2} +\frac{12(2+x)\log x}{5(1-x)^3(x-y)} - \frac{12(2+y)\log y}{5(1-y)^3(x-y)} ~.
\end{eqnarray}

The loop function entering the threshold corrections to the muon mass reads

\begin{equation}
g(x,y) = \frac{2x\log x}{(1-x)(y-x)} - \frac{2y\log y}{(1-y)(y-x)} ~.
\end{equation}


\bibliographystyle{utphys}
\bibliography{FlavorFullSUSY.bib}

\end{document}